\documentclass[12pt, draftclsnofoot, onecolumn]{IEEEtran}
\usepackage{amsmath,epsfig,epsf,times,amssymb,booktabs,subfigure}
\usepackage{setspace}
\usepackage{array}
\usepackage{url}
\usepackage{float}
\usepackage{color}
\usepackage{graphicx}
\usepackage{amsthm,cite}
\usepackage{enumerate}
\usepackage{multirow}

\theoremstyle{remark}

\begin{document}

\title{Intelligent Reflecting Surface for Wireless Communication Security and Privacy}
\author{Shihao Yan, \IEEEmembership{Member, IEEE,} Xiaobo Zhou, \IEEEmembership{Member, IEEE,}\\ Derrick Wing Kwan Ng, \IEEEmembership{Fellow, IEEE}
Jinhong Yuan, \IEEEmembership{Fellow, IEEE,}\\ and Naofal Al-Dhahir, \IEEEmembership{Fellow, IEEE}
\thanks{Shihao Yan, Derrick Wing Kwan Ng, and Jinhong Yuan are with the School of Electrical Engineering and
Telecommunications, The University of New South Wales, Sydney, NSW 2052,
Australia (e-mails: \{shihao.yan, w.k.ng, j.yuan\}@unsw.edu.au).}
\thanks{Xiaobo Zhou is with the School of Physics and Electronic Engineering, Fuyang Normal University, Fuyang, 236037, China (e-mail: zxb@fynu.edu.cn).}
\thanks{Naofal Al-Dhahir is with the Department of Electrical and Computer
Engineering, The University of Texas at Dallas, Richardson, TX 75080, USA
(e-mail: aldhahir@utdallas.edu).}
\thanks{This work has been submitted to the IEEE for possible publication. Copyright may be transferred without notice, after which this version may no longer be accessible.}
}

\maketitle

\vspace{-1cm}

\begin{abstract}
Intelligent reflection surface (IRS) is emerging as a promising technique for future wireless communications. Considering its excellent capability in customizing the channel conditions via energy-focusing and energy-nulling, it is an ideal technique for enhancing wireless communication security and privacy, through the theories of physical layer security and covert communications, respectively. In this article, we first present some results on applying IRS to improve the average secrecy rate in wiretap channels, to enable perfect communication covertness, and to deliberately create extra randomness in wireless propagations for hiding active wireless transmissions. Then, we identify multiple challenges for future research to fully unlock the benefits offered by IRS in the context of physical layer security and covert communications. With the aid of extensive numerical studies, we demonstrate the necessity of designing the amplitudes of the IRS elements in wireless communications with the consideration of security and privacy, where the optimal values are not always $1$ as commonly adopted in the literature. Furthermore, we reveal the tradeoff between the achievable secrecy performance and the estimation accuracy of the IRS's channel state information (CSI) at both the legitimate and malicious users, which presents the fundamental resource allocation challenge in the context of IRS-aided physical layer security. Finally, a passive channel estimation methodology exploiting deep neural networks and scene images is discussed as a potential solution to enabling CSI availability without utilizing resource-hungry pilots. This methodology serves as a visible pathway to significantly improving the covert communication rate in IRS-aided wireless networks.

\end{abstract}

\section{Introduction}

As the fifth-generation (5G) cellular networks are being deployed, industry and academic attention turned into the research and deployment of new technologies for establishing the upcoming sixth-generation (6G) networks.  A fully-connected digital world, e.g., the true realization of the Internet-of-Things (IoT) paradigm, will be potentially enabled by the 6G networks \cite{Giordani2020toward}, in addition to realizing enhanced mobile broadband (eMBB), ultra-reliable and low latency communications (URLLC), and massive machine type communication (mMTC) as promised in 5G networks. Meanwhile, wireless communication security and privacy issues are of ever-increasing importance and are major barriers to the wide deployment of IoT devices in our daily life and thus limit the world economic growth offered by IoT. Also, artificial intelligence (AI) is expected to be heavily integrated into 6G networks in order to reap the benefits brought by machine learning in wireless networks \cite{Giordani2020toward}. In the context of AI, the provisioning of security and privacy become even more critical, due to the wide spread of large amounts of private data and the remote control of machines. Therefore, an enabling technology of 6G networks must be capable of addressing both the security and privacy concerns, while having the ability to pave the way for the deployment of the fully-connected digital world.

Intelligent Reflecting Surface (IRS), which is one disruptive enabler of 6G networks, can smartly reconfigure the wireless propagation environment with high flexibilities \cite{wu2020towards}. The key advantages of the IRS technology are: 1) IRS is highly compatible coexisting with other novel physical layer wireless technologies, as it targets the wireless medium of signal propagation while the other techniques are mainly implemented at the transceivers; 2) IRS consists of a large number of reflecting elements integrated on a planar surface, which normally comprises three layers, i.e., the meta-atom, control, and gateway layers \cite{sena2020NOMA}. As such, IRS is inherently of low hardware complexity and enjoys a high flexibility for practical deployment, e.g., the planar surface can be mounted on a wall or a roof with arbitrary sizes and shapes; 3) In general, the IRS's reflecting elements are passive without the need for power-hungry active radio-frequency (RF) chains and its control components are also ultra-low-power electronic circuits \cite{wu2020towards}. Therefore, IRS is an energy-efficient technology, which can be potentially powered by wireless energy harvesting.
Besides, IRS's benefits in terms of improving wireless communication performance have been widely demonstrated in the literature and some prototyping efforts (e.g., \cite{wu2020towards,sena2020NOMA}). For example, IRS has been used for efficiently mitigating interference to support air-ground communications via passive beamforming to maximize the performance of non-orthogonal multiple access (NOMA) \cite{sena2020NOMA}.

The benefits of IRS in terms of improving wireless communication security have been quantified based on the theories on physical layer security (e.g., \cite{cui2020secureviaIRS,guan2020ANusefulness,yu2020jsacIRS,hong2020ANIRS}).
Indeed, physical layer security can guarantee security of communication contents from an information-theoretic point of view. It is mainly based on advanced signal processing to manipulate the air interface in wireless communications and it does not require encryption or decryption with keys. Thus, it is of low complexity and, for example, can be used to distribute private keys during the network initialization stage. The low-complexity property of physical layer security is synergistic with the low-complexity hardware of IRS, making their integration as an ideal technique for guaranteeing a certain level of secrecy in numerous low-complexity or dynamic wireless networks (e.g., IoT, vehicular networks). In addition, their integration is also desirable in various URLLC application scenarios, as their low complexities do not cost much signal processing time and thus can help reduce communication latency. In particular, IRS is able to significantly enhance physical layer security by controlling its reconfigurable reflecting elements to add wireless signals constructively at a legitimate receiver, but destructively at a potential eavesdropper.

On top of security, IRS has also been identified as an excellent technique to preserve a user's privacy in wireless communications (e.g., \cite{lu2020IRScovert,zhou2020IRSdelay,lv2020IRSnoma,si2020IRSassisted,wang2021IRSactive}). This is achieved through an emerging technology termed covert communication or low probability of detection communication \cite{yan2019lowpro}. This technology aims at hiding the very existence of a wireless transmission from a warden and thus is able to achieve a high-level of privacy (e.g., avoiding exposing a transmitter, or hiding a transmitter's location). IRS can adjust the phase shifts or amplitudes of its reflecting elements to deliberately create a spatial null surrounding the warden and thus achieve high communication covertness. Besides, IRS's potential in achieving covert communications is more distinguishable in indoor scenarios. This is due to the fact that the indoor propagation environment is relatively static with limited random interference and thus the probability of successfully detecting a wireless transmission is generally high. Therefore, the indoor covert communication rate is significantly low and becomes a bottleneck for the entire covert wireless networks. In addition to the signal null-space creation, dynamically varying the reflecting behaviors of IRS's elements introduces more uncertainties in the channels to reduce the detection accuracy at the warden and thus, in turn, significantly enhances the performance of covert communications.

\begin{figure}[!t]
    \begin{center}
        \includegraphics[width=3.8in]{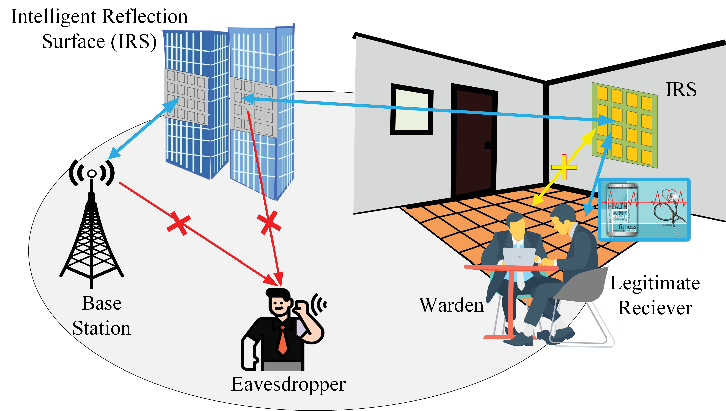}
        \caption{A typical application scenario of adopting IRS to enhance wireless communication security and privacy, where IRS helps protecting communication content (e.g., private medical and health information) against an eavesdropper and aids in preserving a user's privacy by hiding the very existence of a wireless transmission from a warden.}
        \label{fig:scenario}
    \end{center}
\end{figure}

A typical scenario of applying IRS to enhance wireless communication security and privacy is shown in Fig.~\ref{fig:scenario}, where an IRS-aided physical layer security solution can protect the communication contents (e.g., private medical and health data) against an eavesdropper and an IRS-based covert communication solution is able to hide the very existence of a wireless transmission to preserve a user's privacy against a warden. Although some applications of IRS to achieve wireless communication security and privacy have been identified in preliminary works (e.g., \cite{cui2020secureviaIRS,guan2020ANusefulness,yu2020jsacIRS,hong2020ANIRS,lu2020IRScovert,zhou2020IRSdelay,lv2020IRSnoma,si2020IRSassisted,wang2021IRSactive}), many challenges still exist in applying IRS to realize the physical layer security and covert communications in practical scenarios. These challenges include, but are not limited to, how to optimally adjust the reflecting amplitudes of IRS elements subject to security or covertness constraints, how to obtain the IRS's channel state information (CSI) securely or covertly, and how the emerging powerful machine learning paradigm can help to realize the benefits offered by IRS in practical scenarios. Against this background, this article first summarizes the main existing results on IRS-aided physical layer security in Section II. Then, the benefits of IRS in improving covert communications are discussed in Section III. We note that multiple potential solutions to the aforementioned challenges are also presented in these two sections. Furthermore, Section IV discusses how a deep neural network (DNN) helps to achieve accurate CSI of IRS links in practical application scenarios. Finally, some concluding remarks are drawn in Section V.

\section{IRS-aided Physical Layer Security}

In this section, we first present some existing results on IRS-aided physical layer security and then discuss some associated challenges in this direction for future works. We note that the channel estimation challenges are separately discussed in Section IV, since they exist in both the IRS-aided physical layer security and covert communications.

\subsection{Main Existing Results}

\subsubsection{Achieving Non-Zero Secrecy Capacity and Improving Average Secrecy Rate}

The benefits of adopting IRS to enhance the physical layer security was first demonstrated by considering a challenging scenario, where the quality of the main channel from a transmitter, Alice, to a legitimate receiver, Bob, is lower than that of the eavesdropper's channel from Alice to an eavesdropper, Eve \cite{cui2020secureviaIRS}. In such a scenario without the aid of an IRS, a non-zero secrecy capacity is not achievable if each of the legitimate transceivers is equipped with a single-antenna, since the main channel capacity $C_b$ is lower than the eavesdropper's channel capacity $C_e$. We note that the secrecy rate is defined as the maximum value between $0$ and $C_b - C_e$. Following  \cite{cui2020secureviaIRS}, the effectiveness of artificial noise (AN) in IRS-aided systems for guaranteeing the physical layer security was investigated in \cite{guan2020ANusefulness} by considering multiple eavesdroppers. In particular, the authors of \cite{guan2020ANusefulness} showed that the secrecy performance of an IRS-aided system without AN can be worse than that of an AN-aided system without IRS, especially when the number of eavesdroppers near the IRS increases.

\begin{figure}[!t]
    \begin{center}
        \includegraphics[width=3.8in, height=3in]{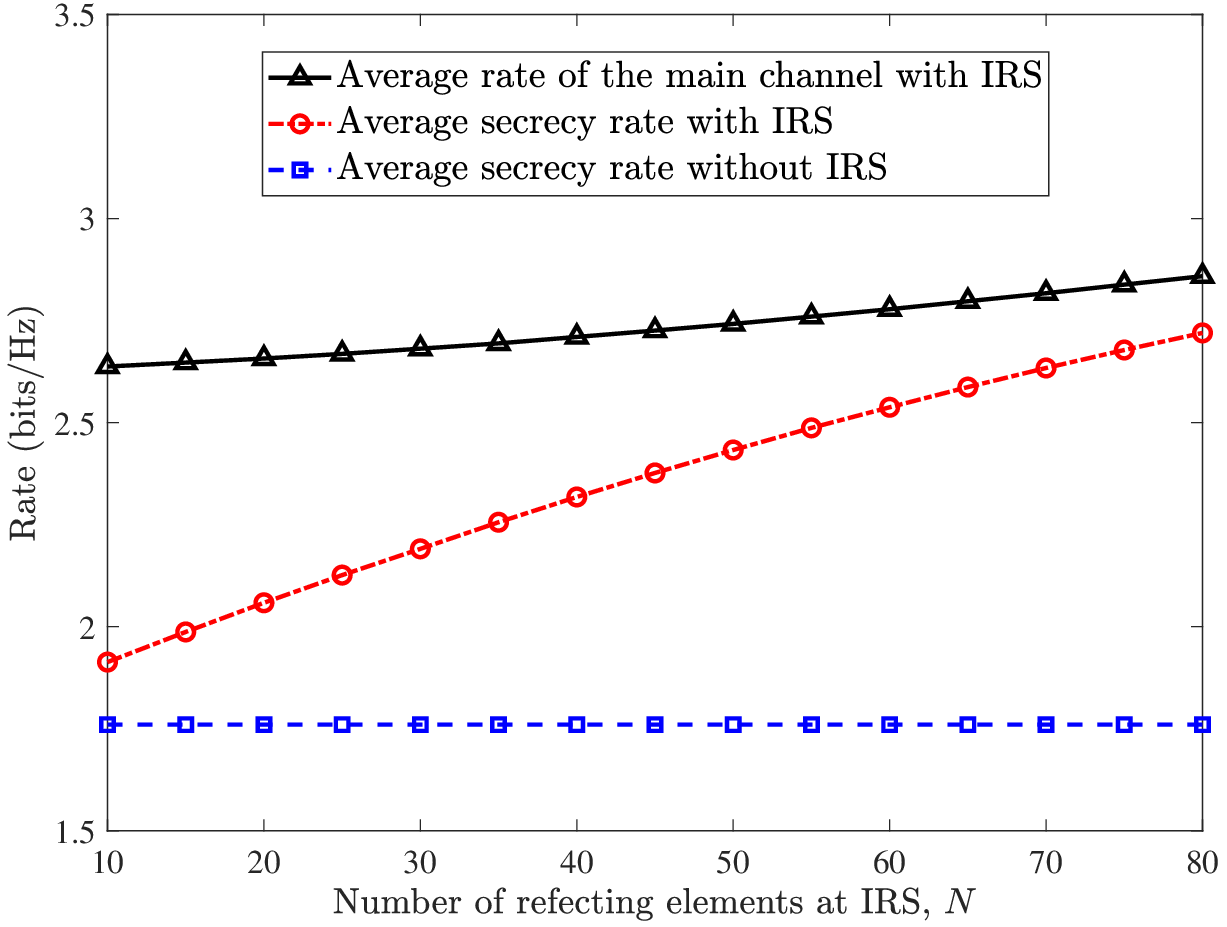}
        \caption{The average secrecy rate of the wiretap channel and the average rate of the main channel versus the number of IRS's elements, where a single-antenna Alice transmits to a single-antenna Bob with the aid of an IRS in the presence of a single-antenna Eve and the CSI of all the channels are available at all nodes for simplicity. All the channels are subject to Rician fading with a Rician factor of $2$ dB. Alice, Bob, Eve, and IRS are respectively located at  $[0,5]$, $[35,10]$, $[75,10]$, and $[55,0]$, where the distance is in meters (m).}
        \label{fig:average}
    \end{center}
\end{figure}

To further demonstrate the benefits of using IRS to enhance the physical layer security, we plot the average secrecy rate, which is obtained by averaging the secrecy rate over multiple channel realizations, versus the number of an IRS's elements in Fig.~\ref{fig:average}. To this end, the IRS's phase shifts are optimized to maximize the secrecy rate for each channel realization according to \cite{cui2020secureviaIRS}. As expected, in this figure we observe that the average secrecy rate of the system with an IRS is significantly higher than that of the system without an IRS and the performance gain increases with the number of the IRS's elements (i.e., $N$). The main reason is that the IRS's ability for focusing information signals at Bob and suppressing them at Eve increases with $N$. In addition, we also plot the average rate of the main channel without considering secrecy in Fig.~\ref{fig:average}, where the gap between it and the average secrecy rate quantifies the cost of guaranteeing the physical layer security in the considered system. Interestingly, in this figure we observe that this gap decreases with $N$, which shows that the IRS's capability of achieving secrecy improves with its number of elements. This reveals a key advantage of IRS in guaranteeing physical layer security.

\subsubsection{Distributed IRSs and Their Locations}

The aforementioned systems assume a single legitimate receiver and an IRS-aided system with multiple receivers was examined in \cite{yu2020jsacIRS}, where multiple distributed IRSs were considered to help the secure transmission from a multi-antenna access point to these multiple single-antenna receivers. It was shown that deploying two IRSs at different locations, is beneficial relative to deploying only one IRS at a fixed location, when the total number of IRS elements is fixed. The main reason is that multiple IRSs at various locations can introduce multiple independent propagation paths. This increases the macro diversity to be exploited and thus improves the possibility in establishing strong end-to-end line-of-sight channels from the access point to the served multiple users. This result demonstrates the importance of optimizing the IRS's location for unlocking the potential of IRS-aided physical layer security.

\subsubsection{Discrete Phase Shifts and Reflection Amplitudes}
Considering multiple antennas at both the legitimate transceivers and the eavesdropper, the authors of \cite{hong2020ANIRS} tackled the secrecy rate maximization problem by jointly designing the transmit beamforming and the IRS's phase shifts. The impacts of discrete phase shifts and reflection element amplitude variation were also numerically examined in \cite{hong2020ANIRS}, which demonstrated that, as expected, the secrecy rate increases with the resolution for each phase shifter. Interestingly, it also showed that when all the reflection amplitudes assume a common value, the secrecy rate increases with this common value. However, we note that this conclusion may not hold if these reflection amplitudes have different values, which will be further discussed in the following subsection.

\subsection{Existing Challenges and Future Works}

\subsubsection{Reflection Amplitude Optimization}

Although the reflection amplitudes $\beta_n$ in IRS can take any value between zero and one, i.e., $\beta_n \in [0,1]$, in most existing works they are set as $\beta_n =1$ to maximize the reflecting signal strength (e.g., \cite{cui2020secureviaIRS,guan2020ANusefulness,yu2020jsacIRS,hong2020ANIRS}). However, this is generally not the optimal setting in some application scenarios requiring the physical layer security, especially when the CSI of the eavesdropper's channel is not available due to the existence of passive eavesdroppers. The main reason is that increasing the reflection amplitudes may also inevitably improve Eve's channel quality on average (when Eve's CSI is unavailable), in addition to increasing Bob's channel quality. Then, considering a constraint on Eve's average channel quality, which is determined by the maximum allowable secrecy outage probability in the considered system, setting all the reflection amplitudes to $1$ is strictly suboptimal in terms of maximizing the secrecy performance. Therefore, identifying the conditions, under which $\beta_n =1$ is not optimal, and tackling the optimization of $\beta_n$, remain unsolved in the context of IRS-aided wiretap channels.

\subsubsection{Number and Location Optimization of Distributed IRSs}

One attribute of IRS is its flexibility to be mounted on a surface with different shapes. In order to fully utilize this attribute for maximizing the benefits brought by IRS, the number and locations of IRSs should be jointly optimized, especially when the total number of IRS reflecting elements is fixed. Although this statement is supported by some preliminary simulation results, e.g., in \cite{yu2020jsacIRS}, a mathematical framework for analysis is still missing in the literature. In fact, this optimization is even more challenging in physical layer security due to, for example, the incomplete information on the number or the locations of the eavesdroppers. In addition, once the number and locations of the distributed IRSs are determined, it would become costly to update them dynamically or frequently. As such, multiple practical factors, e.g., quality of service requirements, hardware limitations, and maintenance accessibility, should be involved in the optimal design of the number and locations of multiple distributed IRSs.

\section{IRS-Aided Covert Communications}

In this section, we first discuss some preliminary results on IRS-aided covert communications, which aim at preserving a user's privacy in wireless communications. Then, we point out some critical research problems that deserve future research efforts along this direction.

\subsection{Main Existing Results}

\subsubsection{Achieving Perfect Communication Covertness}
In covert communications, a transmitter, Alice, intends to transmit information signals to a legitimate receiver, Bob, while maintaining a negligible probability of this transmission being detected by a warden, Willie \cite{yan2019lowpro}. Then, perfect covertness is achieved when Alice communicates with Bob with a non-zero transmit power, while Willie cannot determine whether this transmission occurs or not from his observation (i.e., Willie's detection performance is the same as a random guess). In a single-antenna system without IRS, perfect communication covertness is not achievable, since Willie would definitely receive related measurements as long as Alice transmits information signals. This makes Willie's detection performance better than a random guess. Considering an IRS's capability of signal suppression, with the aid of an IRS, perfect covertness becomes achievable in such a single-antenna system \cite{zhou2020IRSdelay}. Intuitively, this is due to the fact that an IRS can effectively control the Alice-IRS-Willie channel by tuning its phase shifts and amplitudes to enable the signals from the Alice-IRS-Willie channel and the Alice-Willie channel to counteract each other at Willie. As proved in \cite{zhou2020IRSdelay}, the condition for achieving perfect covertness is that the quality of the effective Alice-IRS-Willie channel is at least as good as that of the Alice-Willie channel, due to the fact that the IRS's reflecting elements are passive and thus the maximum value of each IRS's element amplitude is $1$.

\begin{figure}[!t]
    \begin{center}
        \includegraphics[width=3.8in, height=3in]{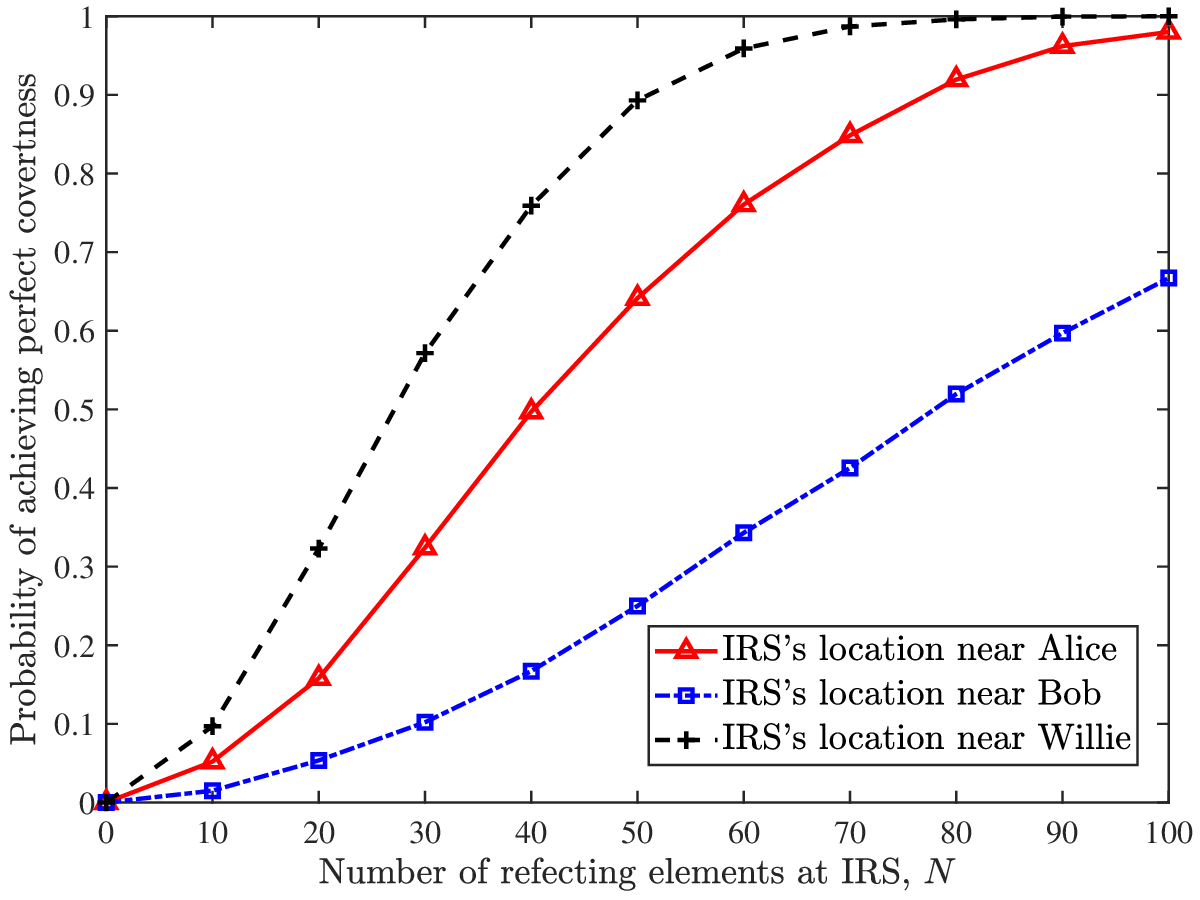}
        \caption{The probability of achieving perfect covertness versus the IRS's number of reflecting elements, where Alice, Bob, and Willie are located at $[0,5]$, $[80,0]$, and $[100,0]$, respectively, while one IRS is put near one of them in three cases, respectively, i.e., the three possible locations of the IRS are $[0,10]$, $[80,10]$, and $[100,10]$. All the channels are subject to Rician fading with a Rician factor of $5$ dB.}
        \label{fig:perfect}
    \end{center}
\end{figure}

To demonstrate the above benefits offered by IRS,  we plot the probability of achieving perfect covertness in such a single-antenna system in Fig.~\ref{fig:perfect} by assuming Rician fading in all involved channels. As expected, in Fig.~\ref{fig:perfect} we first observe that this probability increases with the number of reflecting elements $N$ at the IRS and it approaches to $1$ when $N$ is greater than a certain value. Intuitively, this is due to the fact that the average gain of the Alice-IRS-Willie channel increases with $N$ and thus the probability to have the Alice-IRS-Willie channel being as good as the Alice-Willie channel increases with $N$. In addition, we can observe that the IRS's location also significantly affects the probability of achieving perfect communication covertness. This is mainly due to the considered production path loss model, which makes the effective path loss of the Alice-IRS-Willie channel increase as the IRS moves towards the center between Alice and Willie.


\subsubsection{Creating Extra Uncertainties to Improve Communication Covertness}

Relative to the perfect covertness requirement, a relatively relaxed covertness constraint is that the warden Willie's detection error probability is equal to or larger than a certain value $1-\epsilon$, where $\epsilon$ determines the required covertness level (i.e., the required covertness level decreases with $\epsilon$). In some application scenarios, the warden Willie may know the CSI of the channel from Alice to himself, which enhances his detection of Alice's potential transmission and thus makes it hard to guarantee the covertness constraint. In this context, IRS was proposed for introducing artificial uncertainties at Willie in order to enhance covert communication performance \cite{lv2020IRSnoma}. This is a viable approach since it is hard for Willie to aquire the CSI of the extra Alice-IRS-Willie channel brought by the IRS. As shown in  \cite{lv2020IRSnoma}, through generating a random phase shift pattern at IRS to deliberately insert uncertainties in IRS links, Willie's detection error probability is notably boosted and this probability increases with the number of IRS elements.

\subsubsection{Extending Covert Communication Range}

To satisfy the covertness constraint, the transmit power in covert communications is normally in the low regime, which significantly limits the covert communication range. In order to overcome this limitation, the authors of \cite{wang2021IRSactive} proposed to apply an IRS to extend the communication range from a multi-antenna transmitter Alice to a single-antenna Bob. Specifically, the transmit beamforming and the IRS's phase shifts were jointly optimized to maximize the communication quality from Alice to Bob subject to a covertness constraint. This work also demonstrated the importance of optimizing the IRS's passive beamforming in order to realize the promised performance enhancement by the IRS.

\subsection{Existing Challenges and Future Works}

\subsubsection{Reflection Amplitude Optimization}

\begin{figure}[!t]
    \begin{center}
        \includegraphics[width=3.8in, height=3in]{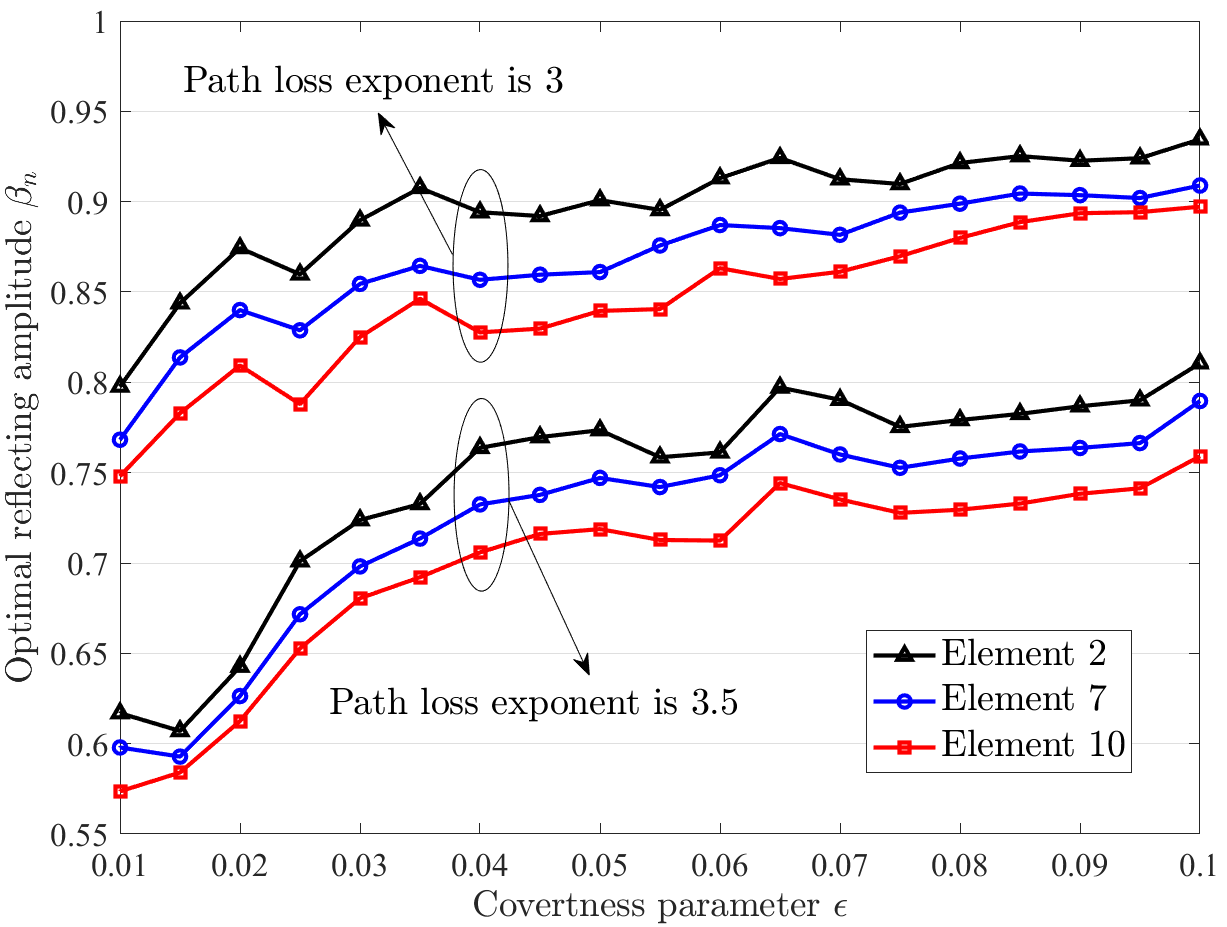}
        \caption{The IRS's optimal reflecting amplitudes for covert communications versus the covertness parameter $\epsilon$, where the covertness constraint becomes more stringent as $\epsilon$ decreases and the IRS's amplitudes together with its phase shifts and the transmit power are jointly optimized via a brute-force search. All the channels are subject to Rician fading with a Rician factor of $3$ dB. Alice, Bob, Willie, and IRS are respectively located at  $[0,5]$, $[35,10]$, $[75,10]$, and $[55,0]$.}
        \label{fig:amplitude}
    \end{center}
\end{figure}

Different from the above positive results of using IRS in covert communications, the work in  \cite{si2020IRSassisted} showed that IRS may be harmful to covert transmissions in some cases, even when the partial or full CSI of Willie's links is available. The main reason is that all the IRS's reflecting amplitudes are set as $1$, which makes the direct and reflected signals add constructively rather than destructively at the warden Willie such that Willie's detection performance is improved. This interesting result reveals the importance of optimally designing the IRS's amplitudes together with the phase shifts in the context of covert communications. For better visualization, we plot the values of the IRS's optimal reflecting amplitudes versus the covertness parameter $\epsilon$ in Fig.~\ref{fig:amplitude}, where we assume that the CSI of the links related to Alice and Bob is available, while the CSI of the links related to Willie is not available. To generate the results presented in this figure, Alice's transmit power is also jointly optimized with the IRS's reflecting amplitudes and phase shifts, in order to minimize the communication outage probability from Alice to Bob for a fixed-rate transmission, subject to a covertness constraint \cite{zhou2020IRSdelay}. As shown in this figure, we find that the IRS's optimal reflecting amplitudes are not necessarily $1$. In fact, the optimal values of the reflecting amplitudes depend on the related channel qualities (e.g., path loss) as shown in Fig.~\ref{fig:amplitude}, which can be far from $1$. Intuitively, this is due to the fact that there are more degrees of freedom in varying multiple amplitudes than varying Alice's transmit power (which is a scalar) in maximizing Bob's receive signal strength subject to the covertness constraint. In addition, in this figure we also observe that these optimal amplitudes generally increase with the covertness parameter $\epsilon$, i.e., their optimal values increase and approach $1$ as the covertness constraint becomes less stringent. This is consistent with the conclusion that setting all the IRS's amplitudes to $1$ is optimal when the covertness constraint is not considered.

\subsubsection{Optimal Detector and Tradeoff between Signal Focusing and Suppression}

For effective covert communication provision, a conservative approach should be adopted, where the best attacking ability of a malicious user should be determined. The determination of an optimal detector in IRS-enabled covert communications is challenging. This is due to the fact that the reflected signals by IRS together with the directly received signals from a transmitter may result in correlated observations at Willie, which can lead to mixed distributions with intractable likelihood functions. In this context, the conventional energy detector may no longer be optimal. Therefore, a key challenge is to develop new detection approaches with proved optimality and easy-to-evaluate performance. From a system design point of view, the IRS has to focus the signals at Bob for improving communication performance, while suppressing the signals at Willie to avoid being detected. Therefore, identifying the optimal trade-off between communication from Alice to Bob and the detection at Willie is another specific technical challenge in IRS-enabled covert communications, which can be potentially resolved based on the min-max theory \cite{yan2019lowpro}. Addressing these two identified challenges is desirable in future works, which would establish a novel foundation for determining the performance limit of IRS-enabled covert communications.

\section{Channel Estimation subject to Security and Privacy Constraints}

In this section, we first discuss the specific challenges and issues for channel estimation in the IRS systems with security and privacy requirements. Then, we propose a potential solution to securely and covertly perform passive channel estimation.

\subsection{Channel Estimation Accuracy Tradeoff and Challenges}

\begin{figure}[!t]
    \begin{center}
        \includegraphics[width=3.8in, height=3in]{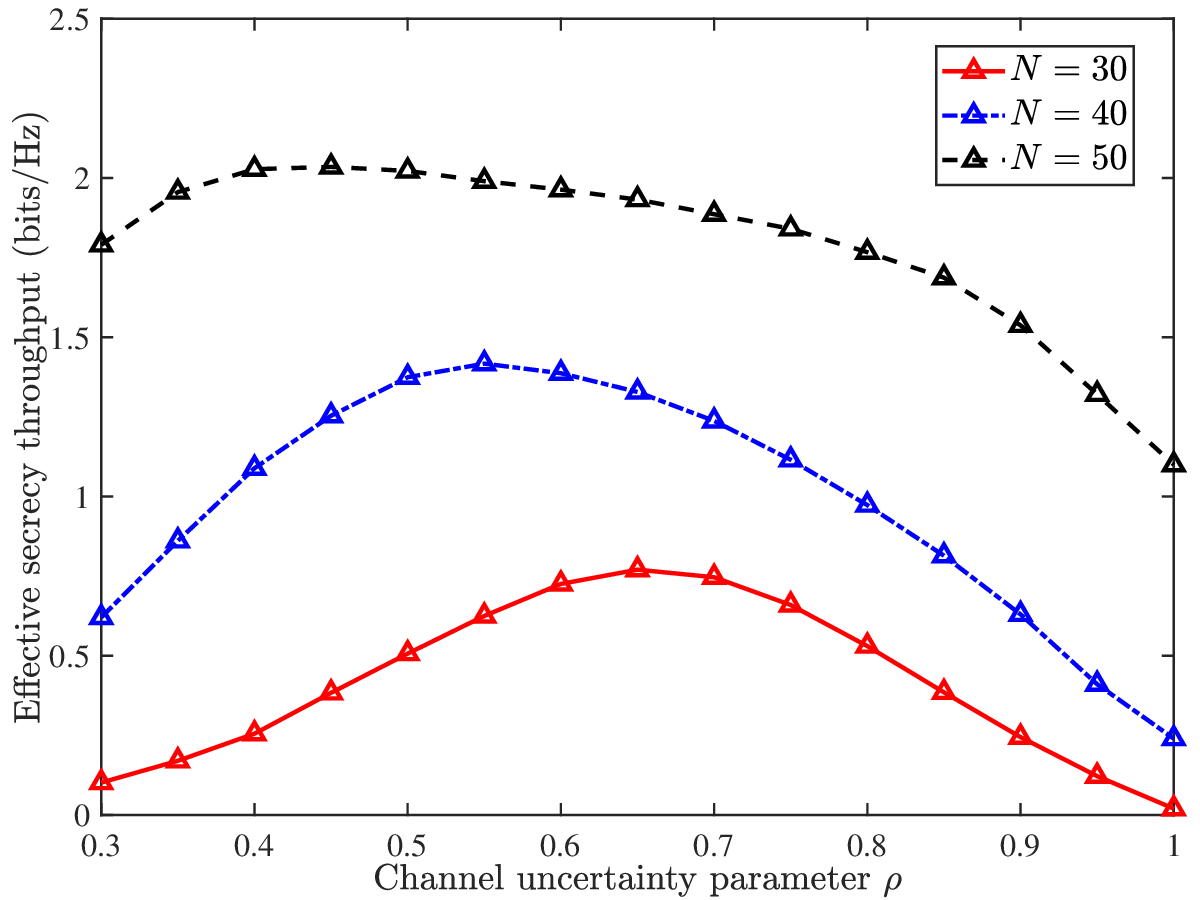}
        \caption{The effective throughput, which is defined as $R_s(1-p_{\text{to}})(1-p_{\text{so}})$, versus the CSI uncertainty parameter $\rho$, where $R_s$ is the secrecy rate, $p_{\text{to}}$ is the transmission outage probability, $p_{\text{so}}$ is the secrecy outage probability, the CSI accuracy increases with $\rho$, e.g., $\rho=0$ means no CSI and $\rho=1$ means perfect CSI. All the channels are subject to Rician fading with a Rician factor of $5$ dB. Alice, Bob, Eve, and IRS are respectively located at  $[0,0]$, $[40,0]$, $[60,0]$, and $[40,10]$.}
        \label{fig:ESTrho}
    \end{center}
\end{figure}

Acquiring accurate CSI is critical in achieving the aforementioned benefits offered by IRS in wireless networks, with or without security and privacy requirements. In fact, channel estimation is a challenging fundamental research problem to the design of IRS-aided wireless communications, due to the significantly increased number of channel coefficients and the lack of active RF chains equipped at IRS. Besides, channel estimation becomes more important in physical layer security. In general, if more resources are allocated to channel estimation to improve the CSI accuracy at the legitimate transceivers (Alice and Bob), the CSI accuracy also increases at the eavesdropper Eve, since Eve is normally assumed to be a powerful receiver such that she knows all the information from a conservative point of view. Then, one key challenge is to strike a balance between channel estimation accuracy and the achievable secrecy level in IRS-based wiretap channels. In order to take the first step to tackle this challenge, we plot the effective secrecy throughput, which quantifies the amount of information bits that can be transmitted from Alice to Bob reliably and securely, versus the CSI uncertainty parameter $\rho$ in Fig.~\ref{fig:ESTrho}. Specifically, the CSI accuracy of the links related to the IRS increases with $\rho$. In this figure, we consider a fixed-rate secure transmission and a single-antenna at each of Alice, Bob, and Eve. The estimated CSI is available at both Bob and Eve, while Alice has only the estimated CSI of the legitimate links due to the considered passive eavesdropping scenario.


From Fig.~\ref{fig:ESTrho}, we observe that the effective secrecy throughput first increases with $\rho$, which is mainly due to the fact that increasing CSI accuracy at Bob improves the transmission reliability. However, as $\rho$ further increases, the effective secrecy throughput decreases, since Eve's decoding capability increases as her CSI accuracy also increases, which leads to a higher secrecy outage probability. If Eve's CSI is publicly available, the result should be different and we conjecture that the effective secrecy throughput will monotonically increase with $\rho$, since Eve's CSI can be utilized by Alice to determine her transmission strategies (e.g., beamforming with or without AN).


Addressing the channel estimation issue in IRS-aided physical layer security calls for new channel training strategies, which will enable only the legitimate transceivers to obtain accurate CSI, but not Eve. If the CSI accuracy increases simultaneously at both the legitimate transceivers and the eavesdropper, as demonstrated in Fig.~\ref{fig:ESTrho}, the amount of resource that is allocated to channel estimation should be carefully optimized. In fact, channel estimation in IRS systems with privacy constraints is more challenging, since, for example, the initial pilot transmission may be detected by the warden Willie and thus violates the user's privacy. Therefore, passive channel estimation is more suitable for acquiring CSI in IRS-based systems with a covertness constraint (i.e., a privacy requirement), relative to the traditional pilot-based channel estimation. One type of such passive channel estimation will be discussed in the following subsection.



\subsection{Passive Channel Estimation based on Machine Learning}

Considering the open nature of the wireless medium, the CSI of a wireless link highly depends on its communication environment, for example, transceivers' location information, surrounding objects' distribution, shapes, or materials. However, the mapping function from this environment information to the CSI, e.g., Maxwell's equation, is complicated and mathematically intractable for real-time resource allocation, which leads to various technical challenges in channel reconstruction based on such information. Recently, owning to the significant advances in machine learning, the determination of CSI from a wireless link's environment information becomes feasible. In particular, the mapping function can be well mimicked by some model fitting adopted in machine learning and these models can be trained accurately with a large amount of real data samples (e.g., \cite{liaskos2020endtoend,xu2021deep}). For example, DNN can be used to extract useful environment information (e.g., relative locations of transceivers and obstacles, blockage materials) for reconstructing CSI from certain scene videos or real-time images. In addition, a geometric model may not be sufficient for mapping such information to CSI, where reinforcement learning can be applied. The determination of CSI from environment information is passive without requiring the transmission of pilots. It provides the potential for keeping all the legitimate users totally silent before any active information transmission and thus can guarantee a high level of secrecy and privacy in wireless communications.


\begin{figure}[!t]
    \begin{center}
        \includegraphics[width=4.8in]{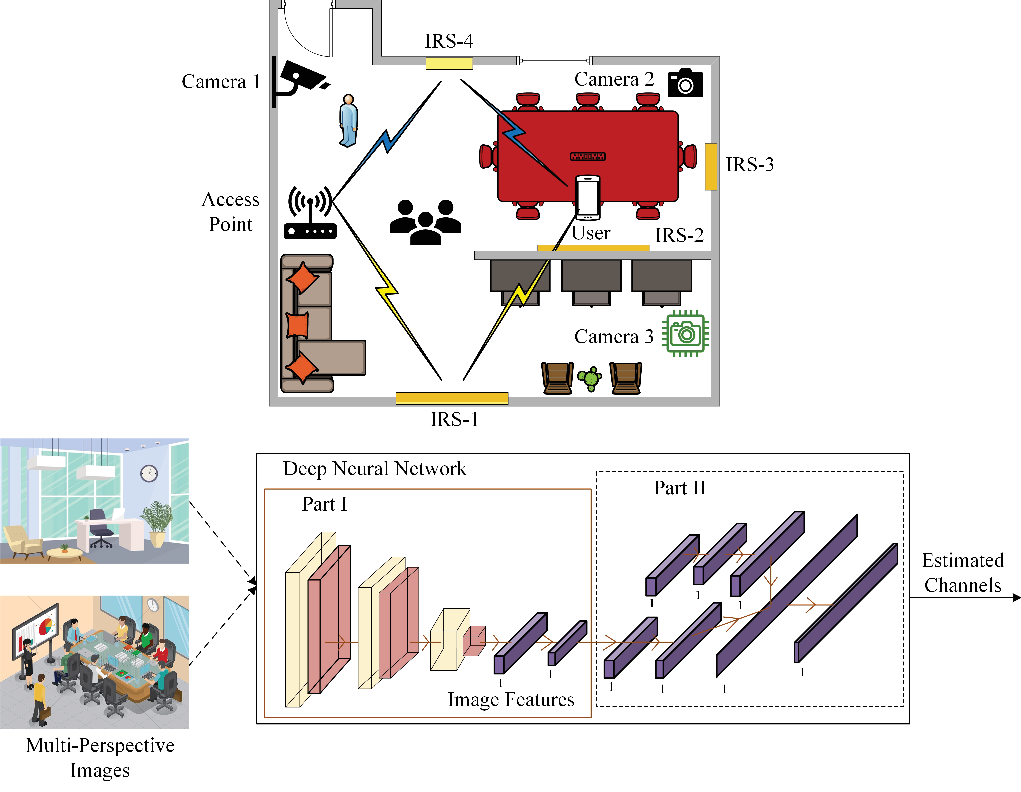}
        \caption{An application scenario of passive channel estimation based on scene images or videos, in order to obtain CSI covertly in IRS-aided covert communication systems. In this scenario, the wireless channel from the access point to the user in the surrounding environment with multiple IRSs can be reconstructed from the images captured by multiple cameras, by using emerging machine learning techniques, where an architecture of a neural network is also presented in this figure.}
        \label{fig:machine}
    \end{center}
\end{figure}

In Fig.~\ref{fig:machine}, we present an application scenario of passive channel estimation based on scene images together with the DNN structure proposed in \cite{xu2021deep}. In this scenario, we aim to estimate the channel between the access point and the user passively, in order to, e.g., avoid potential eavesdroppers to estimate the channel from the access point and prevent potential wardens to discover the existence of the access point or the users. The input of the DNN is a set of images on the communication environment captured by multiple cameras from different angles. The DNN consists of two parts. The first part is to extract the related image features, which determine the properties of the corresponding channels, by adopting multiple convolutional layers. This part can be separately conducted at the user or a single camera in order to facilitate the conveyance of these features to a central unit for obtaining the final channel estimation. We note that this part of the DNN can be accomplished with a federated machine learning structure, where multiple cameras first train the DNN with their own images and then exchange the DNN weights through a central unit. The second part of the DNN is to use all the extracted image features to reconstruct the desired channels at a central unit. If the multiple cameras and the central unit are connected via cables in an application scenario (e.g., an indoor scenario as shown in Fig.~\ref{fig:machine}), the aforementioned passive channel estimation does not require any active signal transmission before the information conveying, which can offer opportunities to hide all the legitimate transceivers and thus establish shadowed and inherently secure wireless networks.




\section{Conclusions}\label{sec:conclusion}

As a potential enabler of 6G wireless networks, IRS has demonstrated its superiorities in different wireless communication scenarios. In this article, we demonstrated its potential in achieving wireless communication security and privacy by summarizing some major benefits brought by IRS into physical layer security and covert communications. Then, multiple research challenges and future works were identified in order to fully unlock the benefits of IRS in guaranteeing communication secrecy and covertness. Finally, special attention was paid to channel estimation issues in IRS-aided systems with security and privacy requirements, and a passive channel estimation based on machine learning was proposed as a potential solution to resolving these issues.

\bibliographystyle{IEEEtran}

\end{document}